# Observable consequences of a hypothetical transient deviation from Quantum Mechanics

Alejandro A. Hnilo.
*CEILAP, Centro de Investigaciones en Láseres y Aplicaciones; CITEDEF, J.B. de La Salle 4397, (1603) Villa Martelli, Argentina..*

December 20th, 2012.

**Abstract:** The conflict between Quantum Mechanics (QM) and the intuitive concepts of Locality and Realism (LR) is manifest in the correlation between measurements performed in remote regions of a spatially spread entangled state. In this paper, it is hypothesized that transient deviations (from the values predicted by QM) occur if the correlation is measured in a time shorter than $L/c$, where $L$ is the spatial spread of the entangled state and $c$ is the speed of light. In this way, the conflict is solved by changing QM minimally. Under general assumptions, it is obtained a mathematical model of the process that reproduces the QM value after a time longer than $L/c$ has elapsed. One of the predictions of this model is that oscillations of the rate of coincidences should exist, with a main frequency lower than $c/4L$. An experiment able to reveal these oscillations is shown to be accessible, by placing stations at $L \approx 5$ Km and reaching a coincidence rate of $\approx 3 \times 10^5$ s$^{-1}$ (a value already obtained at the laboratory scale). This means a test of QM vs LR of a completely new type, with several practical and theoretical advantages.

*Note: this is the complete version of the communication with the same title presented at the conference "Quantum Optics VI" held in Piriápolis, Uruguay, 12-15 November, 2012.*

PACS: 42.50.Xa, 03.65.Ud, 03.65.Ta.

Keywords: Optical tests of Quantum Mechanics, Bell's inequalities, Entanglement and non-locality.

## 1. Introduction.

The Copenhagen interpretation of Quantum Mechanics (QM) has faced debate since its early years. A.Einstein objected the non-determinism intrinsic in the theory, and claimed that it was incomplete. One generation later, J.S.Bell showed that QM is in contradiction with intuitive notions about the locality of physical phenomena, or with the existence of a Universe whose properties are independent of the observer. These notions usually receive the shorthand of Local Realism (LR) [1,2]. J.S.Bell also showed that experiments with entangled states of two particles can decide whether QM or LR is valid in the Nature. The resolution of this controversy is essential to the foundations of Physics. In the proposed experiments, the statistical correlation between the results of measurements performed on the two entangled particles is obtained. LR imposes a limit on this correlation, while QM predicts a correlation higher than that limit, regardless of the distance between the particles. Many versions of these experiments have been performed, generally confirming the violation of the averaged correlation limit imposed by LR.

Nevertheless, giving up LR is so contrary to the common sense that the search for alternative explanations is, at some extent, defensible. Many alternative theories holding to LR, or LR theories (also named "hidden variable theories"), have been proposed to reproduce the observed results. These theories usually exploit the imperfections of the experiments, through one or more of the so-called *logical loopholes*. In general, the proposed LR theories are not claimed to describe a real physical situation. The effort is focused in showing that they are indistinguishable from QM in the practice, and hence, that LR has not been disproved yet. Despite the technical improvements achieved after many years in the measurement of the violation of the correlation limit, some loophole always survives, and the controversy remains undecided. It has even been speculated that, for deep fundamental reasons, the controversy is intrinsically not decidable [3].

It is pertinent to note that most of the proposed LR theories, if they were demonstrated true, would force a complete rewriting of the currently accepted description of the world at the microscopic scale, and to fully discard QM. The overwhelming success of QM in many different physical problems suggests that it is improbable that it ever fails in the prediction of statistically averaged values. Little attention has been devoted instead to the possibility of deviations *in time* (i.e., transient deviations) from the QM predictions, what is known as *non-ergodic* (LR) theories.

This paper explores this possibility, and searches its observable consequences. A model is proposed (it is named "Transient Quantum Mechanics", TQM), by following a path somehow inverse to most LR theories. It does not assume a microscopic world completely different from QM. Instead, it modifies QM as little as possible. Besides, no effort is made to obtain results indistinguishable from the QM predictions. On the contrary, I look for the deviations from QM that arise naturally from the general properties of the model, and that can be effectively observed. An observable deviation is found indeed: quasi-periodical oscillations of the rate of coincidences after the analyzers (see Fig.1) with a main frequency $< c/4L$. The observation of these oscillations implies a test of a completely new type, independent of the violation of a statistical correlation, and with several practical and theoretical advantages. Besides, TQM is not a fully arbitrary manufacture, but it is inferred from a scenario or framework that is (in my opinion) physically plausible.

That framework is named here "Non-ergodic contextuality" (NEC). NEC is not indispensable to define and work with TQM, so that its details are left for the Appendix A. It is anyway convenient mentioning here that NEC assumes that the results of the observations at the quantum scale are not random but determined, in the statistical average at least, by the states of the atoms in the setup and its environment (i.e., *contextuality*). Besides, these atoms evolve in time (i.e., *non-ergodicity*) in a coupled and *chaotic* way. As a result of these features, NEC (and hence TQM) is not ruled out by the impossibility theorems or by the experiments performed until now (Appendix A).

**2. Two hypotheses.**

In this Section, I state and discuss the two hypotheses on which TQM is based.

The conflict QM vs LR arises mainly from the correlation, higher than allowed by LR, between the results of observations performed in remote regions of a spatially spread entangled state. In order to solve the conflict with a minimal modification of QM, it is proposed the following *First hypothesis*:

*H1: The correlation between the results of the observations performed in remote regions of a spatially extended entangled state of size L needs a time >L/c (after an "unpredictable change" in the setup, see the Second Hypothesis below) to reach the value predicted by QM. If the correlation is measured in a time <L/c, then the value obtained holds to LR.*

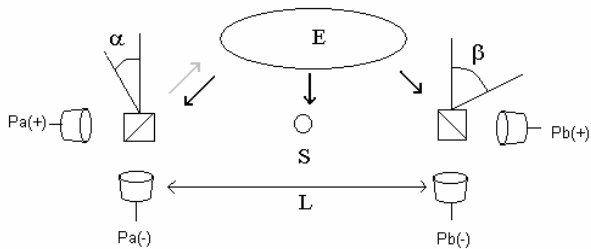

Figure 1: An EPRB setup. The source S emits pairs of photons towards stations placed at a distance $L$. The angle settings of the analyzers are changed in a way *unpredictable by the system* while the photons are in flight. In the TQM model, such unpredictable change in the value of $\alpha$ triggers forces acting only on the environment $E$ (grey arrow). The reaction forces of $E$ (black arrows) are described by Lindblad operators $L^{(i)}$ acting on the field. The spatial spread is taken into account by making the $L^{(i)}$ to act on the state the field had a time $\tau \approx L/c$ before the change. The field is always in a LR-limited state.

It is evident that H1 solves the paradoxes related with the apparent "spooky action at a distance" of entangled states. H1 means that entangled states of arbitrary large size $L$ do not exist, although thinking in terms of such entangled states is a convenient approximation to calculate average rates. The approximation is valid if the time required by the measurement of the correlation is much longer than $L/c$. It is speculated that this is valid to all scales. For example: the correlation between particles on Earth and the Moon would not show entanglement unless the measuring process took longer than $\approx 1$ sec, but entanglement at the atomic scale could be observed after $\approx 10^{-19}$ sec.

H1 is not refuted by any of the experiments performed to date (see Sections 4 and 5). H1 is the only change assumed from usual QM in this paper. All the concepts and the tools of QM remain valid, as far as they are not in conflict with H1. It is worth stating here that H1 has negligible impact on Quantum Computation, because quantum computers will be small and the algorithms will run during a time much longer than $L/c$. In the application known as Quantum Key Distribution, instead, H1 may open a window to an eavesdropper in the case the flux of pairs is larger than $c/L$ (see Sections 4 and 5).

In what follows, because of its relevance and simplicity, I will focus on the case of two photons entangled in polarization: the Einstein-Podolsky-Rosen-Bohm setup (EPRB, see Fig.1). In this case, H1 can be read as a sort of semi-classical hypothesis: the matter is quantized, but the field is classical. "Classical" here does not mean that photons do not exist, but that the photon states hold to LR. In other words: the photon states are separable, non-entangled or, in general, "LR-limited". I will call "system" = setup + environment $E$. The "setup" = source + analyzers + detectors.

An important requirement of the experiments aimed to test QM vs LR is that the values of the angles $\{\alpha,\beta\}$ of the analyzers in Fig.1 must be unpredictable. This is to close the so-called *locality* or *contextual* loophole [4,5]. A key detail is that the values must be unpredictable *by the system*. This leads to several questions: even if $\{\alpha,\beta\}$ are changed during the flight of the photons from the source to the detectors, how to be sure that the system cannot predict, at least partially, that change? And why leaving $\{\alpha,\beta\}$ fixed should be less predictable *by the system* than changing them? Closing this loophole is difficult, for there can always be some hidden correlation between the source of the changes and the system [6]. This possibility cannot be simply excluded, even if only partial contextuality is suspected to be valid in the Nature, because the degree of statistical correlation sufficient for a contextual LR theory to reproduce the QM values is surprisingly low [6-8]. There is no reliable way to ensure that the settings are unpredictable *enough*. The importance of this problem has been pointed out by several authors [8-10]. It is an intricate issue that has even metaphysical consequences, leading to debates on the existence of free will [11] or a strictly deterministic Universe [12]. From the point of view of an experimentalist, the problem seems a dead end impossible to escape without making *at least one* unverifiable and controversial assumption. The NEC framework shows a way out that is, in my opinion, reasonable. The basic idea arises from the observation that, in some time scale, all physical systems become unpredictable.

In NEC, the coupled evolution of the different parts of the system is the cause of the high statistical correlation observed. But, the exponential increase of the distance between initially neighbor points in phase space, which is characteristic of chaotic dynamics, eventually destroys that coupled evolution. These spontaneous "losses of track" are equivalent, from the point of view of the system, to an unpredictable change of the settings. It is then natural, inside the NEC picture, to propose the following *Second hypothesis*:

*H2: Changes unpredictable by the system occur spontaneously (at an average rate $\mu \neq 0$).*

Between the two extremes (absolute free will or strict determinism) H2 means an intermediate position: any physical system evolves deterministically during periods, which are interrupted by unpredictable events. The origin of the unpredictability would be then physical, not metaphysical, and besides, it would be unavoidable. The precise value of $\mu$ would be an attribute of the system's dynamics, probably related with the value of its highest Lyapunov exponent [13]. An important practical consequence of H2 is that events equivalent to unpredictable changes should occur even if the analyzers' settings in the Fig.1 are kept fixed.

It is evident that the decisive way to test whether H1 is valid in the Nature is to measure the violation of the LR correlation limit in a time shorter than $L/c$ after an unpredictable change. As it is shown later, this is an impossible task nowadays. In order to find an alternative, accessible test, the gap from the general idea to a mathematical model providing definite numerical predictions must be filled. In the next Section, some assumptions are made, in compliance with H1 but as unspecific and simple as possible, to get that mathematical model.

### 3. Transient Quantum Mechanics (TQM).

Let assume that the source (in Fig.1) emits pairs of photons in the fully symmetrical Bell state: $|\varphi^+\rangle = (1/\sqrt{2}) \{|x_a, x_b\rangle + |y_a, y_b\rangle\}$. If observations on $a$ and $b$ are performed, the correlation between the results, as predicted by QM, is higher than allowed by LR (see Fig.2). In the experiments, this correlation is often quantified with the Clauser-Horne-Shimony and Holt parameter $S_{CHSH}$ [1]. It involves the measurement of the probability of double coincidences at the "transmitted" ports of the analyzers $P^{++}(\alpha,\beta)$, at the "reflected" ports $P^{--}(\alpha,\beta)$, and the mixed cases $P^{+-}(\alpha,\beta)$ and $P^{-+}(\alpha,\beta)$, at the angle settings $\alpha=\{0, \pi/4\}$ and $\beta=\{\pi/8, 3\pi/8\}$. The QM value is $S_{CHSH} = 2\sqrt{2}$, while LR imposes that $S_{CHSH} \leq 2$.

The TQM model is built in three steps:

*3.1 First step: LR-limited state of the field.*

According to H1, there are no spatially spread entangled states. Hence, the spatially spread state of the field in the Fig.1 must be LR-limited. This state must also be able to display the QM correlation, at least in some condition. In order to fulfill the two requirements, note that the probability values predicted by QM can be reproduced, regardless the value of $\beta$, by a mixture of photon pairs polarized parallel and orthogonal to the axis $\alpha$, or:

$$\rho_\alpha = \tfrac{1}{2} \{|\alpha\rangle\langle\alpha|_a \otimes |\alpha\rangle\langle\alpha|_b + |\alpha_\perp\rangle\langle\alpha_\perp|_a \otimes |\alpha_\perp\rangle\langle\alpha_\perp|_b\} \quad (1)$$

this state is classical, separable, or LR-limited. It is able to reproduce the QM predictions only because the value of $\alpha$ is known. When $\alpha=0$ ($\pi/4$), $\rho_{\alpha=\pi/4}$ ($\rho_{\alpha=0}$) generates no correlations, i.e., $P^{ij}(\alpha,\beta)= \tfrac{1}{4} \;\forall i,j,\alpha,\beta$. Thus, it also reproduces the observable effect of the unpredictable change (or the spontaneous loss of track) discussed in the previous Section.

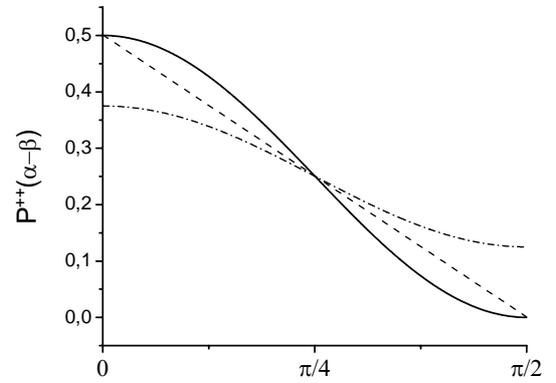

Figure 2: The higher-than-classical correlation between remote regions of the entangled state $|\varphi^+\rangle$ is illustrated by this plot of the probability of coincidence as a function of the difference angle between the analyzers in the Fig.1. Dotted straight line: limit imposed by LR, no curve can be above this line between $[0,\pi/4]$ and below it between $[\pi/4,\pi/2]$; solid curve: QM prediction, $P^{++}= \tfrac{1}{2}\cos^2(\alpha-\beta)$; point-dot curve: a semi-classical theory, $P^{++}= \tfrac{1}{4}[\tfrac{1}{2} + \cos^2(\alpha-\beta)]$.

*3.2 Second step: Lindblad approximation.*

The unpredictable change in $\alpha$ assumed in H2 triggers forces of reaction inside the system. The evolution that follows is complex. In order to describe this evolution without making any specific assumption on the composition of the system and the involved interactions, only the observable effects can be taken into account. I assume then that the forces of reaction acting on the field come only from $E$, and describe their effect with the Lindblad operator:

$$d\rho_\alpha/dt = L_\mu \rho_\alpha L^\dagger_\mu - \tfrac{1}{2} L^\dagger_\mu L_\mu \rho_\alpha - \tfrac{1}{2} \rho_\alpha L^\dagger_\mu L_\mu \quad (2)$$

where the $L_\mu$ describe the action of $E$ on $\rho_\alpha$, as it is often done to describe decoherence. The Lindblad approximation means that the reaction forces inside $E$ dissipate in a negligible short time. The eq.(2) is hence valid only in a certain "coarse" time scale and for a Markovian $E$ [14]. Let take:

$$L^{(1)} = g\ (-|x_a,x_b\rangle+|y_a,y_b\rangle)(\langle x_a,y_b|+\langle y_a,x_b|) \quad (3)$$

where the parameter $g$ measures the strength of the interaction. Its value is unknown. From the NEC picture, it is conceivably related with the distribution, density, composition and temperature of the atoms in $E$. The operator $L^{(1)}$ makes the LR-limited state $\rho_{\alpha=\pi/4}$, which does not reproduce the QM predictions for $\alpha=0$, to evolve into the LR-limited state $\rho_{\alpha=0}$, which does (see Appendix B). The operator that produces the other required evolution, $\rho_{\alpha=0} \to \rho_{\alpha=\pi/4}$ is:

$$L^{(2)} = g\ (|x_a,y_b\rangle+|y_a,x_b\rangle)(\langle x_a,x_b|-\langle y_a,y_b|) = -L^{(1)\ t} \quad (4)$$

The appropriate operator is $L^{(1)}$ or $L^{(2)}$ depending on the value of $\alpha$. The interaction described by the eqs.(2-4) fulfills the condition of being unspecific, for it is defined by its observable effects only. No attempt is made to describe its details or mechanisms, to keep it as general as possible.

The probabilities of observing a coincidence are computed as $\text{Tr}[\rho_\alpha.Q_\alpha^{(a)}\otimes Q_\beta^{(b)}]$, being the $Q_\alpha^{(a)}$ the operator of projection for the passage through an analyzer oriented at an angle $\alpha$ and acting on the subspace $a$. Thanks to the form of $\rho_\alpha$ and the $Q$'s [1] the following equalities hold $\forall \alpha,\beta$:

$$P^{++}(\alpha,\beta) = P^{--}(\alpha,\beta),\ P^{+-}(\alpha,\beta) = P^{-+}(\alpha,\beta) \quad (5)$$

$$P^{++}(\alpha,\beta) + P^{+-}(\alpha,\beta) = \tfrac{1}{2} \quad (6)$$

The probabilities can be then written in terms of a single parameter $\rho_d$:

$$P^{++}(\alpha,\beta,t) = \cos^2\alpha.\{\rho_d(t).\cos^2\beta + \tfrac{1}{2}.[1-2.\rho_d(t)]\sin^2\beta\} +$$
$$+ 2.\cos\alpha.\sin\alpha.\cos\beta.\sin\beta.\{1-2.\rho_d(t)\} +$$
$$+ \sin^2\alpha.\{\rho_d(t) + \tfrac{1}{2}.[1-4.\rho_d(t)].\cos^2\beta\} \quad (7)$$

where (see Appendix B):

$$d\rho_d(t)/dt = -4g^2\ [\rho_d(t) - \rho_{\text{target}}] \quad (8)$$

where $\rho_{\text{target}} = \tfrac{1}{4}$ (or $\tfrac{1}{2}$) if the setting is $\alpha=\pi/4$ (or $\alpha=0$).

As an illustration, let assume an initial condition corresponding to $\alpha=\pi/4$ and that an unpredictable change to $\alpha=0$ occurs at $t=0$. The probability of a double-passage coincidence evolves then as:

$$P^{++}(\alpha=0,\beta,t) = \tfrac{1}{2}\cos^2\beta + \tfrac{1}{4}\ exp(-4g^2 t).(1-2.\cos^2\beta) \quad (9)$$

that is, $P^{++}(\alpha=0,\beta,t)$ goes from the uncorrelated value $\tfrac{1}{4}$ at $t=0$ to the QM prediction as $t\to\infty$. The operator $L^{(2)}$ produces the analog evolution when the unpredictable change is from $\alpha=0$ to $\alpha=\pi/4$. In this way, the QM predictions are reproduced by a LR-limited state for $t\to\infty$. This is, precisely, what is assumed by H1.

Note that the matrices obtained from tracing out one photon: $\text{Tr}[\rho_\alpha]_{\text{partial}} = \tfrac{1}{2}\mathbf{I}\ \forall t$, so that no time evolution is observed in the detection of single photons.

*3.3 Third step: the state of the field is delayed.*

The forces represented by the $L^{(j)}$ act over a spatially spread region. The information on the value of $\alpha$ (which defines which one of the $L^{(j)}$ is acting) needs a time $>L/c$ to reach the other side of $E$. The consequence is that the reaction to an unpredictable change must be delayed a time $>L/c$. This is taken into account in the model by using the value of $\rho_\alpha$ as it was a time $\tau \approx L/c$ before the change. In other words: in the statistical average, the state of the field is delayed a time $\tau$ with respect to the instantaneous value of $\alpha$. This delay is the formal way to introduce non-ergodicity (i.e., dependence on the history of the system, see Appendix A). The eq.(8) then becomes:

$$d\rho_d(t)/dt = -4g^2\ [\rho_d(t-\tau) - \rho_{\text{target}}(t)] \quad (10)$$

where the function $\rho_{\text{target}}(t)$ is externally defined (it represents the "unpredictable changes"), $\rho_{\text{target}} = \tfrac{1}{2}$ ($\tfrac{1}{4}$) if $\alpha=0$ ($\pi/4$). The eq.(10) completes the mathematical formulation of TQM. Note that the usual QM result is retrieved in the limit $\tau\to 0$, $g^2\to\infty$.

The eq.(10) is a delay differential equation. It evolves in a phase space of infinite dimensions. This is a convenient property, for it is intended to embed the high dimensional dynamics implicit in NEC. On the other hand, the features of its solutions are difficult to find in the general case. A glimpse, valid if $\mu\tau<<1$ (i.e., low rate of unpredictable changes) is obtained by studying the case $\rho_{\text{target}}(t) = $ constant. In this case, eq.(10) can be written:

$$dx(\theta)/d\theta = -\Gamma\ x(\theta-1) \quad (11)$$

where $x(\theta) \equiv \rho_d(\theta) - \rho_{\text{target}}$, $\theta \equiv t/\tau$, and $\Gamma = 4g^2\tau$.

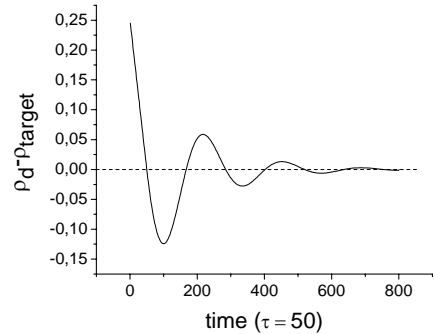

Figure 3: solution of eq.(11) for $\Gamma=1$ and initial condition $x(t<0) = \tfrac{1}{4}$. It describes the evolution following an unpredictable change, from $\alpha=0$ to $\alpha=\pi/4$, occurring at $t=0$.

In the fig.3, the solution of eq.(11) is plotted if $x(t<0) = \tfrac{1}{4}$, $\Gamma=1$ and an unpredictable change occurs at $t=0$. The change is followed by a damped quasi-periodical oscillation towards the target state, with decay time $\approx 3.5\tau$ and period $\approx 4.5\tau$. The physical cause of the oscillations is that the action of $E$ into the field

"overshoots" because of the delay. Enforcing locality (of the propagation of the information on the value of $\alpha$ by making $\tau \neq 0$) is therefore sufficient to obtain the oscillations.

In the fig.4, the variation of the decay time and the (approximate) period of the solutions of eq.(11) are plotted as functions of $\Gamma$. The amplitude of the oscillations diverges if $\Gamma > \pi/2$. Note that for $\Gamma > 0.6$ the period is nearly constant, and equal to $\approx 4.5\tau$. This is an important result regarding an experimental test.

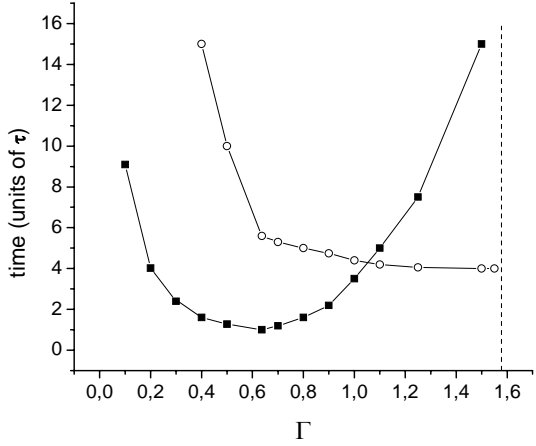

Figure 4: Values of the decay time (full squares) and the period of oscillation (open circles) in units of $\tau$, for the solutions of eq.(11) with initial condition $x(t<0)=\frac{1}{4}$. The vertical dotted line indicates the value of $\Gamma$ at which the amplitude of the oscillations diverge.

**4. Testing TQM.**

A numerical simulation of an EPRB experiment with unpredictable changes in the analyzers' settings is performed. The evolution of $\rho_d(t)$ is calculated from eq.(10), where the value of $\rho_{target}(t)$ is changed in the following way: think a coin is tossed at random times, at an average rate $\mu$. Depending if the result is head or tail, the setting is adjusted to $\alpha=0$ (then $\rho_{target}= \frac{1}{2}$) or $\alpha=\pi/4$ (then $\rho_{target}= \frac{1}{4}$) with a negligible short time of transition. As a result, $\rho_{target}(t)$ jumps between the values $\frac{1}{2}$ and $\frac{1}{4}$, and $\rho_d(t)$ follows it after some delay and damped oscillations (see Fig.5).

The observable, averaged value of $S_{CHSH}$ is obtained by numerical integration. Using the eqs.(5-7), it is:

$$S_{CHSH} = \frac{8\sqrt{2}}{T} \int_0^T \left| \rho_d(t) - \rho_{no-t\arg et}(t) \right| dt \qquad (12)$$

where $\rho_{no-target}(t)$ is $\frac{1}{4}$ ($\frac{1}{2}$) if $\alpha=0$ ($\pi/4$). Note that the QM value $2\sqrt{2}$ is retrieved if $\rho_d = \rho_{target} \; \forall t$.

For fixed $\Gamma$, $S_{CHSH}$ fits the QM prediction if $\mu\tau<1$, and (as it can be expected) it falls below 2 as $\mu\tau>>1$. Yet, the result $S_{CHSH}= 2\sqrt{2}$ can be obtained for any value of $\mu\tau$, by tuning the (unknown) value of $\Gamma$. F.ex., for $\mu\tau=13$ (the apparent value in the experiment in [5]), $S_{CHSH}=2.828$ for $\Gamma=1.549$. The eq.(12) assumes that the rate of detected pairs is arbitrarily high, so that the details of the evolution of $\rho_d(t)$ are exactly followed. A more realistic simulation, where that rate is only $10^{-2}$ or $10^{-3} \tau^{-1}$, provides a poorer discrimination between TQM and QM. In summary: a procedure different from the measurement of $S_{CHSH}$ is necessary to test TQM.

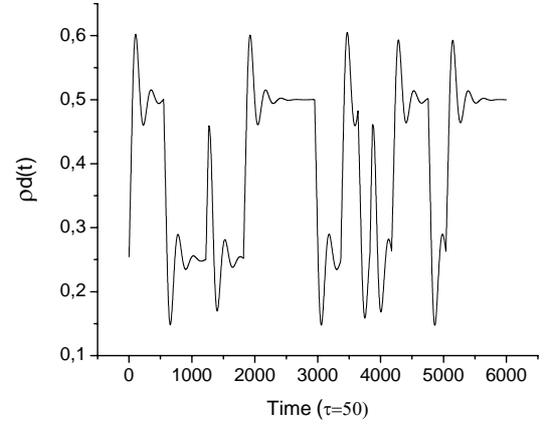

Figure 5: Zoom of the evolution of $\rho_d(t)$, $\Gamma=0.9$, $\mu\tau=0.2$; note the jumps between the two target values, $\frac{1}{4}$ (that corresponds to $\alpha=\pi/4$) and $\frac{1}{2}$ (that corresponds to $\alpha=0$); $S_{CHSH} = 2.79$ for the complete file lasting $2000\tau$.

Note now that for the parameters' values in the fig.5 ($\mu\tau=0.2$ and $\Gamma=0.9$), $S_{CHSH}= 2.79$, too close to $2\sqrt{2}$ to allow a discrimination between QM and TQM in the practice, but that the damped oscillations produce a broad, yet clearly visible, peak in the FFT power spectrum at the period $\approx 4.5\tau$ (fig.6). In the case that $\mu\tau>>1$ (i.e., highly unpredictable settings) and $\Gamma$ tuned so that $S_{CHSH} \approx 2\sqrt{2}$, the peak is sharper and higher than in the case $\mu\tau<1$, but it remains in the same position. It only shifts, towards *lower* frequencies, if $\Gamma<0.6$.

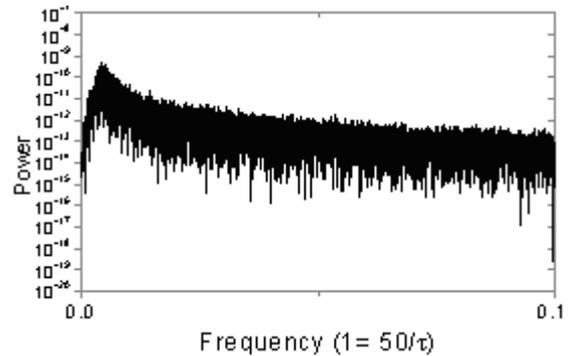

Figure 6: Power FFT spectrum of the complete file of Fig.5, note the peak at $0.213\tau^{-1}$ or period $\approx 4.5\tau$, almost two orders of magnitude above a noisy background.

The accessible test that was looked for is evident now: looking for an oscillation of period $>4L/c$. As $P^{++}(\alpha,\beta)$ is linearly dependent of $\rho_d(t)$ (eq.7), the oscillation can be detected in the number of coincidences after the analyzers. This is a test completely different from all the ones attempted until now. It is independent of the violation of a statistical correlation limit. The precise value of $\mu$ is irrelevant: it suffices that $\mu\neq 0$ (H2). The (often cumbersome) evaluation of the number of accidental counts has negligible impact, and hence, the experiment is also more robust against noise than measuring $S_{CHSH}$. The so-called coincidence-loophole, if it existed, would involve time shifts of the order of the time coincidence window, too short to appreciably distort the oscillations (see Appendix A). A definite result of the test is obtained even if the detectors are not perfect and event-ready signals [1] are not available: if oscillations (with a period varying linearly with $L$) are observed, the usual form of QM is disproved. If the oscillations are not observed instead, TQM is disproved.

In the next Section, the conditions for an experimental test of TQM are discussed. But, before going on: the violation of the LR limit $S_{CHSH} \leq 2$ obtained even when $\mu\tau \gg 1$ is possibly surprising, so that it deserves some comment. When $\Gamma$ is tuned close to $\pi/2$ (i.e., the point where the amplitude of the oscillations of eq.11 diverges), the $P^{ij}(\alpha,\beta)$ obtained from eq.(10) can take, transitorily, values outside the [0,1] interval. It is well known that extended probability values allow a LR theory to reproduce the QM predictions, although it is arguable that such situation actually holds to LR [15]. A possible interpretation for the appearance of probabilities outside [0,1] is, precisely, that non-stochastic processes (as the ones involved in any of the loopholes) are active. Anyway, the important result at this point is this: regardless of the particular mechanism, "conspiracy" or loophole the LR model exploits to fit the QM results in the average, a peak at frequency $\approx c/4L$ (or lower, depending of the unknown value of $\Gamma$) is present.

## 5. Experimental requirements for the test.

In order to reveal the oscillations predicted by TQM, it is necessary to record the time value at which each photon is detected. This technique is named *time stamping* or *time-tag*. It implies some instrumental complication, but it is anyway unavoidable to close the coincidence-loophole [16]. Revealing a frequency peak by random sampling a noisy series is an involved issue. In order to get an idea of the situation, the usual criterion of recording at least two samples in the period of interest suffices. A sample of the value of $P^{++}(\alpha,\beta,t)$ requires the detection of least 10 coincidences. Therefore, one needs 20 pairs after the analyzers in a period $\approx 4.5\tau$ (the lowest value for the period), or $\approx 5$ pairs in $\tau$. In what follows, I consider only the most unfavorable situation, $\tau = L/c$.

Let see now how far the performed experiments are from the goal of a rate of 5 detected pairs in $\tau$. The highest reported rate of entangled pairs was obtained by matching the pumped volume in the non-linear crystal (used to generate parametric fluorescence) with the mode of collecting single-mode optical fibers [17], reaching $\approx 3\times 10^5$ s$^{-1}$ in a laboratory environment. This number cannot be improved, for the best currently available single photon detectors (avalanche photodiodes) cannot be used reliably with a rate detection above $10^6$ s$^{-1}$. Therefore, in order to detect 5 pairs (with the rate reported in [17]) in a time $L/c$, $L>5$ Km. Detection of pairs at 13 Km [18] and even 144 Km [19] has been achieved, but the coincidence rate was much lower than needed, between 50 and 8 s$^{-1}$. Nevertheless, these experiments were performed under the unfavorable conditions of free air propagation. The numbers would improve if the propagation is through a controlled environment or optical fibers. Of course, translating the brightness and purity of a source tested in the lab to a field installation of a size of several Km is a difficult challenge, but it does not seem impossible. A practical advantage is that it is not necessary to achieve $S_{CHSH} \approx 2\sqrt{2}$ to detect the oscillations. Even a poor value of the average correlation would suffice. If a test were attempted by measuring $S_{CHSH}$ in $t<\tau$ instead, $L$ scales to 1000 Km, and a good value of the state's purity should be achieved.

It must be mentioned that dynamics faster than the "coarse" time imposed by the Lindblad approximation may produce peaks at higher frequencies. They cannot be predicted by the TQM model, but of course they may exist in the experiment. More important from the point of view of the test, peaks at lower frequencies may also exist, caused by "revivals" of the coherence in a realistic, imperfectly Markovian $E$. Besides, the numbers discussed above are for the most unfavorable value of the unknown parameter $\Gamma$. As a consequence of all this, the criterion of 5 coincidences in $L/c$ is the most stringent condition to observe the oscillations. A lower value may suffice. An early search performed on the time-stamped data of the experiment in [5] revealed no oscillations [20], but the rate in that experiment was only $2\times 10^{-3}$ $\tau^{-1}$ [21]. The rate of detected pairs should be thus increased a factor about $10^3$ with respect to the currently achieved values (in EPRB setups with remote stations) to enter the range where the oscillations can be expected to be detectable. As it was stated, this is difficult, but not unattainable.

## 6. Summary.

A hypothesis (H1) that reconciles QM with LR is proposed. It minimally changes QM. From H1 and the Lindblad approximation, a simple and unspecific model (TQM) is deduced, which has the mathematical form of a delay differential equation. If the Nature is not strictly deterministic (i.e., if H2 is valid) TQM predicts the existence of quasi-periodic fluctuations of the rate of coincidences detected after the analyzers in an EPRB setup.

That prediction leads to a test of QM vs LR of a new type, independent of the violation of a correlation limit. In the usual test, threshold values of the detectors' efficiencies and the unpredictability of the changes, as well as the knowledge of the moment of emission of the pairs, are necessary to close all the known loopholes [1,10,22, Appendix A]. This is difficult to achieve, and the controversy remains undecided. In the test proposed here, instead, a definite answer is obtained even if those threshold values are not reached and the moment of emission is unknown: if oscillations with a period varying linearly with $L$ are observed, the usual form of QM is disproved. If the oscillations are not observed instead, TQM is disproved.

The test should be performed in a large-size ($L \approx 5$ Km) EPRB setup with a rate of pair detections after the analyzers $\approx 5$ $c/L$. This requires an increase $\approx 10^3$ of the currently available rates in large-size EPRB setups, up to the values already reached at the lab scale. The required time stamping resolution is $\approx 10^{-5}$ s, which is trivially achievable. The analyzers can be kept fixed and the average correlation does not need to be high, these being significant practical simplifications.

**Appendix A: Non-ergodic contextuality (NEC).**

*A.1 Description.*

I describe here the physical picture or conceptual framework supporting H1 and H2. Recall that TQM is deduced from H1, H2 and the Lindblad approximation only. I mean: what follows is not necessary to obtain the results presented before.

NEC arises from the following reasoning: one photon is an amount of energy so small, that it is natural that the result of any process involving it (say, the passage through an analyzer) is influenced by the surrounding atoms. The number of these atoms is enormous, but it is conceivable that the result of their influence converges to a value determined by the boundary conditions or symmetries of the whole. Note that the same happens in the process of frequency conversion in a nonlinear crystal.

NEC assumes that the observed correlations are caused by the influence of all the atoms in the system (i.e., the context) and besides, that this influence evolves in time depending of the system's history (i.e., non-ergodicity). Let review briefly the meaning of contextuality and non-ergodicity:

Contextual theories were defined by J.S.Bell: "*The result of an observation may reasonably depend not only on the state of the system (including hidden variables) but also on the complete description of the apparatus [i.e., the context]*". Contextual theories are not ruled out by the impossibility theorems as Gleason's or Kocher and Specker's ones (see [1] p.1924-25). It is precisely this fact what led J.S.Bell to obtain his inequalities. These theories define the *contextuality*, *predictability* or *locality* logical loophole. In short, contextuality means that the usually reliable approximation of isolating a part of the Universe for its study (neglecting the rest) has found its limit in the EPRB experiment.

The non-ergodic theories [22] state that the *ensemble averages* assumed in QM are not equivalent to the *time averages* obtained in an actual experiment. It was imagined that "*a field, medium or ether with a relatively stable states or memory exists*" that influences the probabilities of detection from one particle to the next, and that acts so as to reproduce the QM predictions in the average over long times (in NEC, the role of the ether is played by the system itself). They are usually related with the *coincidence, trapping* or *memory* logical loophole [3,10,23-25].

The coincidence-loophole allows a LR theory to fit the QM predictions by shifting the photons' detections in time, according to the analyzer's angle, in or out of the time window that defines a coincidence. This makes the number of coincidences to depend on the settings of both analyzers, although the entire process is local. The theories exploiting this loophole cannot be tested even in a setup with 100% efficiency of detection and 100% unpredictable settings. Not a single photon is lost: all what happens is that its detection is delayed or advanced. These theories can only be tested in a setup where the possible time of emission of each pair and the time of detection of each single photon are recorded. The simplest way of understanding these requirements is as follows: imagine the photon detection is delayed until the analyzer changes to the position convenient to fit the QM results. If the expected moment of arrival of the photon is unknown, the delay is unnoticed. There are mechanisms more complex than the one just described, but the delays involved in the coincidence-loophole always are, necessarily, a fraction of $\tau$. Hence, they do not distort the oscillations predicted by TQM, which have a period $> 4.5$ $L/c$.

Because of the large number of atoms involved, the evolution assumed by NEC takes place in a phase space of high dimension. Because of the type of interactions involved, the dynamical equations have many crossed, nonlinear terms. The result is a high dimensional chaotic dynamics (sometimes called *hyperchaos*). As the long-range interactions are mostly electro-magnetic, if an unpredictable (or caused externally to the system) change in the distribution of the atoms (or a spontaneous loss of track as well) occurs in a spatially spread setup, the system requires a time $>L/c$ to adjust its evolution to the new distribution of atoms. This is, precisely, H1.

It is reasonable to ask here *how* a LR model can reproduce the observed non-classical correlations after a time $>L/c$. There are several ways to do it. Possibly the first one (historically) is a mechanical model [24]: an array of classical oscillators, linked by nearest neighbor interactions, produced one photon detection when the value of a certain dynamical variable exceeded a threshold. Among the most recent proposals, the *deterministic learning machines* (DLM) [26] assume that the optical elements in the setup act as units following a simple and adaptive program to process information carried by the photons, to decide when, where, or if, the photons are detected. These approaches, as many others alike, are able to reproduce the QM correlations after a time $>L/c$. But, they make specific hypotheses involving the system and the interactions, and the values of several parameters must be adjusted. The generality of their predictions is uncertain. It is desirable, instead, to find the most general possible features that follow from H1. This is the aim of TQM, and the reason why the Lindblad approximation is used.

I have just mentioned examples of how a LR model can reproduce the QM correlation. Next, I present a symmetry argument (i.e., not based in any specific model) aimed to make plausible *why* it does so.

*A.2 A symmetry argument.*

The aim here is to understand why, although entangled states of the field are assumed not to exist, the system evolves to produce the same averaged statistical correlations as if an entangled state of the field were actually present.

In general terms, a correlation is the consequence of some symmetry invariance. Classical correlations are mostly the consequences of the symmetry *action = reaction*. It is therefore convenient to look for the symmetry invariance, additional to the classical ones, whose consequence is the extra correlation observed in the Fig.2. Let consider the general form of the density matrix of two photons. After taking into account general restrictions and assuming rotational symmetry, that general form is ($\rho_d, \rho_a \in \Re$):

$$\rho \equiv \begin{pmatrix} \rho_d & 0 & 0 & \rho_a \\ 0 & \tfrac{1}{2}-\rho_d & 2\rho_d - \rho_a - \tfrac{1}{2} & 0 \\ 0 & 2\rho_d - \rho_a - \tfrac{1}{2} & \tfrac{1}{2} - \rho_d & 0 \\ \rho_a & 0 & 0 & \rho_d \end{pmatrix} \quad (A1)$$

in the basis $\{|x_a,x_b\rangle, |x_a,y_b\rangle, |y_a,x_b\rangle, |y_a,y_b\rangle\}$ from up to down and from left to right. Be aware that $\rho_\alpha$ in eq.(1) does not hold to eq.A1. The matrix $\rho$ is positive iff:

$$\tfrac{1}{2} \pm |2.\rho_d - \tfrac{1}{2}| \geq \rho_d - \rho_a \geq 0 \quad (A2)$$

The probability of double passage is:

$$P^{++}(\alpha,\beta) = \cos^2\alpha.\{\rho_d.\cos^2\beta + \tfrac{1}{2}.[1-2.\rho_d]\sin^2\beta\} +$$
$$+ \cos\alpha.\sin\alpha.\cos\beta.\sin\beta.\{4\rho_d - 1\} +$$
$$+ \sin^2\alpha.\{\rho_d + \tfrac{1}{2}.[1-4.\rho_d].\cos^2\beta\} \quad (A3)$$

(note the difference with eq.7). The state $\rho$ does not necessarily display a non-classical correlation. F.ex., a semi-classical theory of radiation (SCRT), that assumes that the two photons of the pair are emitted with the same well-defined polarization, which randomly changes from one pair to the next, holds to eqs.A1-A2 with values $\rho_d=3/8$, $\rho_a=1/8$. Then $P^{++}(\alpha,\beta)= \tfrac{1}{4}[\tfrac{1}{2} + \cos^2(\alpha-\beta)]$ and $S_{CHSH} = \sqrt{2} < 2$.

Let take now a fresh view to the Fig.1 when $\alpha=\beta$. The setup is left-right symmetrical (also *E*, at least in the average), as in a mirror. Therefore, if one photon is detected at $P_a^{(+)}$, it is natural to expect that the same happens at $P_b^{(+)}$, for it is simply the image in the mirror. If this does not happen, then there must be some cause, different in each side of the setup. Say, a different value of a "hidden" variable invisible in the mirror (f.ex., the well-defined polarization of the photons in the SCRT mentioned above). But if there is no such hidden difference (precisely as it is assumed in QM), the result of each measurement performed at *a* must be identical to the one performed at *b*, and then:

$P^{++}(\alpha=\beta)= \tfrac{1}{2}$. When this condition is imposed to eqs.A1-A3, $\rho_d = \tfrac{1}{2} = \rho_a$ is the only solution, and the eq.A1 becomes the matrix of the state $|\varphi^+\rangle$ [27]. This result is not a mere coincidence. If the setup is not well assembled and the mirror symmetry is not perfect, so that $P^{++}(\alpha=\beta)= \tfrac{1}{2}-\varepsilon$ ($\varepsilon$>0), then the Concurrence of the resulting state is 1-8$\varepsilon$. In other words: the deviation from the mirror symmetry, and the deviation from "perfect entanglement", both are measured by the same number.

In summary: the symmetry of the system implies that the averaged statistical correlation must be the one obtained as if the state $|\varphi^+\rangle$ existed. But the actual state of the field is LR-limited. Hence, if the settings of the analyzers are changed in an unpredictable way, the field is not able to reproduce that correlation. Nevertheless, the system's symmetry will bound the following evolution in such a way (regardless the details) to reach a state of the field able to reproduce that correlation after a transient has elapsed.

*A.3: Final comments.*

The Bell states are pure states, i.e., entities with no internal parts. They are "atoms" in the original (ancient Greek) meaning of the word. That's why the correlations they produce are higher than classically allowed: what is being observed at the remote stations in the Fig.1 is not a pair of related things, but *the same thing* (to stress this idea, D.Klyshko coined the term *biphoton*). NEC states that "atoms" of arbitrary large size do not exist. NEC states that a time >$L/c$ is needed until the system evolves into a state showing statistical correlations equivalent to the presence of a large "atom". The cause the system evolves in this way is its symmetry as a whole. If this symmetry changes, the (averaged) correlation will change in accordance.

QM would be then an approximation, valid in the average over long periods of time, to a very complex process that involves a macroscopic number of *true* atoms and LR-limited fields. A detailed mathematical description of this process is practically impossible. One has to rely on statistics, as in QM. But, there is an important difference: in QM, it is assumed that each consecutive measurement is independently performed on identical copies of the (entangled) state, being the differences among the observed results the consequence of the intrinsic randomness of the microscopic world. In NEC, instead, the differences observed between consecutive measurements are the consequence of the chaotic dynamics underlying. In QM, the statistical correlation is caused by an entity named entangled state. In NEC, the correlation is caused by the average restrictions on that chaotic evolution imposed by the system's symmetry as a whole.

NEC completes QM, as envisioned by Einstein. It can be understood as a form of the old "statistical interpretation" of QM, updated or enlarged with chaotic dynamics and symmetry arguments. But, faced with attaining a mathematical description, NEC is much more difficult to deal with than QM. In the overwhelming number of cases, the QM approximation is preferable because of easier mathematics and accurate averaged results (although it fails to describe the transients). The success of QM would be the consequence of its capacity for taking into account, in a simple and compact way (through the algebra of operators), the symmetries of the system and the measuring process.

**Appendix B: Evolution due to the Lindbladian.**

In this Appendix, the calculations showing that the operators $L^{(1)}$ and $L^{(2)}$ do produce the required evolutions (from a state unable to reproduce the QM correlations, to a state able to do it) are detailed. It is convenient writing the form of the matrix $\rho_\alpha$ explicitly, in the same basis than eq.A1:

$$(1/8)\begin{bmatrix} 3+\cos(4\alpha) & \sin(4\alpha) & \sin(4\alpha) & 1-\cos(4\alpha) \\ \sin(4\alpha) & 1-\cos(4\alpha) & 1-\cos(4\alpha) & -\sin(4\alpha) \\ \sin(4\alpha) & 1-\cos(4\alpha) & 1-\cos(4\alpha) & -\sin(4\alpha) \\ 1-\cos(4\alpha) & -\sin(4\alpha) & -\sin(4\alpha) & 3+\cos(4\alpha) \end{bmatrix} \quad (B1)$$

therefore:

$$\rho_{\alpha=0} = \begin{bmatrix} \frac{1}{2} & 0 & 0 & 0 \\ 0 & 0 & 0 & 0 \\ 0 & 0 & 0 & 0 \\ 0 & 0 & 0 & \frac{1}{2} \end{bmatrix} \quad (B2)$$

and:

$$\rho_{\alpha=\pi/4} = \begin{bmatrix} \frac{1}{4} & 0 & 0 & \frac{1}{4} \\ 0 & \frac{1}{4} & \frac{1}{4} & 0 \\ 0 & \frac{1}{4} & \frac{1}{4} & 0 \\ \frac{1}{4} & 0 & 0 & \frac{1}{4} \end{bmatrix} \quad (B3)$$

Inserting eq.(2) in eq.(1), $d\rho_\alpha/d(g^2t)$ is then (with obvious notation):

$$\begin{pmatrix} \rho_{22}+\rho_{23}+\rho_{32}+\rho_{33} & -(\rho_{12}+\rho_{13}) & -(\rho_{12}+\rho_{13}) & -\rho_{22}-\rho_{23}-\rho_{32}-\rho_{33} \\ -(\rho_{21}+\rho_{31}) & -2\rho_{22}-(\rho_{23}+\rho_{32}) & -2\rho_{23}-(\rho_{22}+\rho_{33}) & -(\rho_{24}+\rho_{34}) \\ -(\rho_{21}+\rho_{31}) & -2\rho_{32}-(\rho_{22}+\rho_{33}) & -2\rho_{33}-(\rho_{23}+\rho_{32}) & -(\rho_{24}+\rho_{34}) \\ -\rho_{22}-\rho_{23}-\rho_{32}-\rho_{33} & -(\rho_{42}+\rho_{43}) & -(\rho_{42}+\rho_{43}) & \rho_{22}+\rho_{23}+\rho_{32}+\rho_{33} \end{pmatrix}$$

(B4)

Note that there are only four types of equations: for the elements in the corners of the main diagonal (called $\rho_d$), in the corners of the anti-diagonal ($\rho_a$), in the central 2x2 block ($\rho_m$) and all the others, whose derivative is proportional to the sum of elements of the same type. As the initial condition of these elements is equal to zero (see eq.B3), they remain zero $\forall t$. The equations for the nonzero elements are then:

$$d\rho_{m,a}/dt = -4g^2\rho_m \quad (B5)$$

$$d\rho_d/dt = 4g^2\rho_m \quad (B6)$$

$$\rho_d(t) + \rho_m(t) = \frac{1}{2}, \; \rho_d(t) + \rho_a(t) = \frac{1}{2}, \; \forall t \quad (B7)$$

If the initial condition is $\rho_{\alpha=\pi/4}$ (eq.B3) the solution for $\rho_d(t)$ is:

$$\rho_d(t) = \frac{1}{4}(2 - exp(-4g^2t)) \quad (B8)$$

so that the state converges to $\rho_{\alpha=0}$ as $t\rightarrow\infty$.

The operator $L^{(2)}$ generates the transition from $\rho_{\alpha=0}$ to $\rho_{\alpha=\pi/4}$. Using now eq.(3) in eq.(1), $d\rho_\alpha/d(g^2t)$ is:

$$\begin{pmatrix} -2\rho_1+\rho_{14}+\rho_{41} & \rho_{42}-\rho_{12} & \rho_{43}-\rho_{13} & \rho_1+\rho_{44}-2\rho_{14} \\ \rho_{24}-\rho_{21} & \rho_1+\rho_{44}-(\rho_{14}+\rho_{41}) & \rho_1+\rho_{44}-(\rho_{14}+\rho_{41}) & \rho_{21}-\rho_{24} \\ \rho_{34}-\rho_{31} & \rho_1+\rho_{44}-(\rho_{14}+\rho_{41}) & \rho_1+\rho_{44}-(\rho_{14}+\rho_{41}) & \rho_{31}-\rho_{34} \\ \rho_1+\rho_{44}-2\rho_{41} & \rho_2-\rho_{42} & \rho_3-\rho_{43} & \rho_4+\rho_{41}-2\rho_{44} \end{pmatrix}$$

(B9)

and the equations for the nonzero elements are now (see eq.B2):

$$d\rho_{m,a}/dt = 2g^2(\rho_d - \rho_a) \quad (B10)$$

$$d\rho_d/dt = -2g^2(\rho_d - \rho_a) \quad (B11)$$

with solution, if the initial condition is $\rho_{\alpha=0}$:

$$\rho_d(t) = \frac{1}{4}(1 + exp(-4g^2t)) \quad (B12)$$

which converges to $\rho_{\alpha=\pi/4}$ as $t\rightarrow\infty$. Note that also in this case $\rho_d(t) + \rho_a(t) = \frac{1}{2} \forall t$.

The $L^{(i)}$ operators produce the required evolutions also if the initial condition is the SCRT state mentioned in the Appendix A and the Fig.2.

**Auxiliary material.**

This text includes:

- Calculation of the general form of the 4x4 density matrix of two photons in an EPRB experiment.
- Calculation of the form of the density matrix if rotational invariance and $P^{++}(\alpha=\beta)= \frac{1}{2}$ hold.
- Calculation of the form of the density matrix for the other states of the Bell's basis.
- Calculation of the Concurrence if $P^{++}(\alpha=\beta)= \frac{1}{2}-\varepsilon$ ($\varepsilon>0$).
- Useful matrices in explicit form.

- <u>Calculation of the general form of the density matrix.</u>

For two photons and their polarization degree of polarization, in explicit form:

$$\rho = \begin{pmatrix} Rxxxx & Rxxxy & Rxyxx & Rxyxy \\ Rxxyx & Rxxyy & Rxyyx & Rxyyy \\ Ryxxx & Ryxxy & Ryyxx & Ryyxy \\ Ryxyx & Ryxyy & Ryyyx & Ryyyy \end{pmatrix} \quad (X1)$$

in the basis $\{|x_a,x_b>,|x_a,y_b>,|y_a,x_b>,|y_a,y_b>\}$ from up to down and from left to right. Therefore, the sub indices $i, j$ of the (complex) numbers *Rijkl* correspond to the photon in the field mode "a", and the $k, l$ to the one in the field mode "b".
Let impose the restrictions:

1) *General properties of a density matrix.*
   a) $\rho$ is self-adjoint
   b) $Tr(\rho) = 1$
   c) $\rho$ is positive.

The first condition means that $Rikjl=(Rkilj)^*$ and that all the diagonal elements are real numbers. The second one, that the sum of the latter is equal to 1. We reserve the third condition for later use (this is just to deal with a simpler algebra).

2) *Properties of the reduced density matrices.*

If observed separately, each mode must be non-polarized. This means that each reduced matrix must be equal to one-half the 2x2 identity matrix, or:

$$Rxxxx + Rxxyy = Ryyxx + Ryyyy = \tfrac{1}{2}, \quad Rxxxx + Ryyxx = Rxxyy + Ryyyy = \tfrac{1}{2} \quad (X2)$$

$$Rxyxx + Rxyyy = Ryxxx + Ryxyy = 0, \quad Rxxxy + Ryyxy = Rxxyx + Ryyyx = 0 \quad (X3)$$

Using eqs.(X2), *Ryyxx=Rxxyy* and *Ryyyy=Rxxxx*. Using also the property 1-b above, $Rxxyy = \frac{1}{2} - Rxxxx$. Finally, using the eqs.(X3) the number of independent elements of the matrix outside the diagonals reduces from 4 to 2.

3) *Interchange of the particle labels a↔b.*

It implies that $Rxyyx = Ryxxy = Rxyyx^*$, so that both are real numbers. Also, that $Rxyxx = Rxxxy$, so that all the elements of matrix outside the diagonals become determined by a single number. Note that, even though the Bell state $|\psi^->=(1/\sqrt{2})\{|x_a,y_b>-|y_a,x_b>\}$ changes sign at this operation, its density matrix remains invariant.

4) *Interchange of the polarization labels x↔y.*

It implies $Rxyxy=Ryxyx=Rxyxy^*$, so that both are real numbers. Also, that $Rxxxy=Ryyyx= -Rxxxy^*$, so that both are pure imaginary numbers. Note that, even though the Bell states $|\varphi^-\rangle = (1/\sqrt{2})\{|x_a,x_b\rangle-|y_a,y_b\rangle\}$ and $|\psi^-\rangle$ change sign at this operation, their density matrices remain invariant.

It is convenient to display the general form of the matrix at this point, as:

$$\rho_{aux} = \begin{pmatrix} d & ig & ig & c \\ -ig & \frac{1}{2}-d & f & -ig \\ -ig & f & \frac{1}{2}-d & -ig \\ c & ig & ig & d \end{pmatrix} \tag{X4}$$

where: $(c, d, f, g)$ are real numbers.

From now on, the symmetries to be added are determined by the characteristics of the source of the radiation (f.ex., in setups that prepare pairs of photons by parametric fluorescence, type I phase matching leads to rotationally symmetrical states, for type II the symmetry of the state depends on the phase between the ordinary and the extraordinary beams). These are data of the specific setup being used. Depending of the particular symmetries, one or another entangled (or classical) state is obtained.

- <u>Calculation of the form of the density matrix if rotational invariance holds.</u>

*5) Rotational invariance*.

This is not valid for all the possible states of two qubits, classical or not. For example, the Bell states $|\varphi^-\rangle = (1/\sqrt{2})\{|x_a,x_b\rangle - |y_a,y_b\rangle\}$ and $|\psi^+\rangle = (1/\sqrt{2})\{|x_a,y_b\rangle + |y_a,x_b\rangle\}$ are not rotationally invariant (they are twist invariant, see later). The condition $\mathbf{R^{-1}}\rho_{aux}\mathbf{R} = \rho_{aux}$ (see the form of $\mathbf{R}$ at the end of this text) implies that $g=0$ and $4d-2c-2f-1 = 0$. Hence, the general form of the rotationally invariant density matrix is:

$$\rho_R \equiv \begin{pmatrix} d & 0 & 0 & c \\ 0 & \frac{1}{2}-d & 2d-c-\frac{1}{2} & 0 \\ 0 & 2d-c-\frac{1}{2} & \frac{1}{2}-d & 0 \\ c & 0 & 0 & d \end{pmatrix} \tag{X5}$$

which is the form in eq.A1 in the paper ($c = \rho_a$, $d = \rho_d$). The restriction 1-c above, which had not been applied yet, is easy to handle now. The eigenvalues of $\rho_R$ are: $\{d - c$ (twice), $c - d + \frac{1}{2} \pm |2d - \frac{1}{2}|\}$ and all of them must be $\geq 0$, so that:

$$\frac{1}{2} \pm |2d - \frac{1}{2}| \geq d - c \geq 0 \tag{X6}$$

which is the eq.A2 in the paper. The eq.A3 in the paper for $\alpha=\beta$ leads to:

$$P^{++}(\alpha=\beta) = d \qquad (\forall \alpha) \tag{X7}$$

Using the "mirror" symmetry, $P^{++}(\alpha=\beta)= \frac{1}{2} \Leftrightarrow d = \frac{1}{2}$, then $c = \frac{1}{2}$ (from X6), and the eq.A1 in the paper becomes the matrix of the state $|\varphi^+\rangle$.

- Calculation of the general form of the density matrix - Other states of the Bell's basis.

If the source emits photons with crossed polarizations (i.e., states with terms of the form $|xy\rangle$) the mirror symmetry obviously changes to $P^{++}(\alpha=\beta)= 0$, then $\rho_d = 0$ (from X7), $\rho_a = 0$ (from X6), and the eq.A1 in the paper becomes the density matrix of the state $|\psi^-\rangle = (1/\sqrt{2})\{|x_a,y_b\rangle - |y_a,x_b\rangle\}$.

The two remaining states of the Bell's basis, $|\varphi^-\rangle$ and $|\psi^+\rangle$, are invariant to a rotation $\theta$ in a-space and $(-\theta)$ in b-space (twist invariance). Starting from (X4), the condition $\mathbf{T^{-1}\rho_{aux}T}= \rho_{aux}$ (see the form of $\mathbf{T}$ at the end of this text) leads to the general form of the twist invariant density matrix :

$$\rho_T \equiv \begin{pmatrix} d & 0 & 0 & c \\ 0 & \frac{1}{2}-d & -2d-c+\frac{1}{2} & 0 \\ 0 & -2d-c+\frac{1}{2} & \frac{1}{2}-d & 0 \\ c & 0 & 0 & d \end{pmatrix} \qquad (X8)$$

and the condition 1-c imposes that:

$$\frac{1}{2} \pm |2d - \frac{1}{2}| \geq d + c \geq 0 \qquad (X9)$$

The expression of the double passage probability is now:

$$P^{++}(\alpha,\beta) = \cos^2\alpha \{d\cos^2\beta + \frac{1}{2}(1-2d)\sin^2\beta\} - \cos\alpha.\sin\alpha.\cos\beta.\sin\beta.\{4d-1\} +$$
$$+ \sin^2\alpha \{d - \frac{1}{2}(4d-1)\cos^2\beta\} \qquad (X10)$$

which is analogous to the eq.A3 in the paper, with the only difference that the middle term has changed sign. Taking into account the twisted symmetry assumed, the (so to say, "twisted") mirror symmetry means that the analyzers must be oriented at opposite angles ($\alpha= -\beta$) to get maximum or minimum coincidences, then:

$$P^{++}(\alpha=-\beta) = d \qquad (\forall \alpha) \qquad (X11)$$

which leads to $d = \frac{1}{2}$ for the case of maximum coincidences. Then, from (X9), $c = -\frac{1}{2}$ and the density matrix of the state $|\varphi^-\rangle$ is obtained. For minimum coincidences instead, $P^{++}(\alpha=-\beta)= 0$, $d=0$ and $c=0$ (from X9). Then the density matrix of the state $|\psi^+\rangle$ is obtained.

- Calculation of the Concurrence.

To confirm the equivalence between "mirror symmetry" and "entanglement", let's see that they are measured by the same number. Note that once $\rho_R$ (or $\rho_T$) is obtained from the classical symmetries, the only effect of the mirror symmetry is to define the value of $d$. Hence, a natural way to measure an imperfection of the mirror symmetry (for the state $|\varphi^+\rangle$, for example) is to use a (small) parameter $\varepsilon$ such that $d = \frac{1}{2} - \varepsilon$ ($\varepsilon>0$). Then, from X6:

$$\tfrac{1}{2} - \varepsilon \le c \le \tfrac{1}{2} - 3\varepsilon \tag{X11}$$

Using the central value $c = \tfrac{1}{2} - 2\varepsilon$ the matrix $\rho_R$ becomes:

$$\rho_\varepsilon \equiv \begin{pmatrix} \tfrac{1}{2}-\varepsilon & 0 & 0 & \tfrac{1}{2}-2\varepsilon \\ 0 & \varepsilon & \varepsilon & 0 \\ 0 & \varepsilon & \varepsilon & 0 \\ \tfrac{1}{2}-2\varepsilon & 0 & 0 & \tfrac{1}{2}-\varepsilon \end{pmatrix} \tag{X12}$$

The magnitude named "Concurrence" is a standard measure of the entanglement of an arbitrary state. For the case of two qubits, it can be calculated as:

$$\text{Concurrence}(\rho) = C(\rho) = \max(0, \lambda_1 - \lambda_2 - \lambda_3 - \lambda_4) \tag{X13}$$

where $\lambda_i$ are the square roots of the eigenvalues of the matrix $\rho.\rho'$ in decreasing order, and where:

$$\rho' = (\sigma_y^a \otimes \sigma_y^b)\, \rho^* \,(\sigma_y^a \otimes \sigma_y^b) \tag{X14}$$

where $\sigma_y^a$ is the spin-flip Pauli matrix acting in the a-subspace. The calculation is simplified by noting that:

$$\rho_\varepsilon = (1-4\varepsilon)|\varphi^+\rangle\langle\varphi^+| + 2\varepsilon|\psi^+\rangle\langle\psi^+| + \varepsilon|x_a,x_b\rangle\langle x_a,x_b| + \varepsilon|y_a,y_b\rangle\langle y_a,y_b| = \rho_\varepsilon^* \tag{X15}$$

the first two terms on the middle side are invariant at the action of the Pauli matrix, and the other two flip into one another, so that $\rho_\varepsilon' = \rho_\varepsilon$. We are then left to calculate the eigenvalues of $\rho_\varepsilon^2$. They are $\{(1-3\varepsilon)^2, 4\varepsilon^2, \varepsilon^2, 0\}$. Therefore:

$$C(\rho_\varepsilon) = 1 - 6\varepsilon \tag{X16}$$

So that the parameter that measures the deviation from the mirror symmetry also measures the defect from maximum entanglement. A different choosing of the value of $c$ in the interval defined by (X11) just leads to a different coefficient of $\varepsilon$ (the limit values are 4 and 8). The choosing $(1-8\varepsilon)$ makes the concurrence of the SCRT equal to zero. The same result is obtained for the other states of the Bell's basis.

- Useful matrices in explicit form.

The 4x4 rotation matrix, an angle $\theta$ in a-subspace and an angle $\varphi$ in b-subspace:

$$\begin{pmatrix} \cos\theta.\cos\varphi & \cos\theta.\sin\varphi & \sin\theta.\cos\varphi & \sin\theta.\sin\varphi \\ -\cos\theta.\sin\varphi & \cos\theta.\cos\varphi & -\sin\theta.\sin\varphi & \sin\theta.\cos\varphi \\ -\sin\theta.\cos\varphi & -\sin\theta.\sin\varphi & \cos\theta.\cos\varphi & \cos\theta.\sin\varphi \\ \sin\theta.\sin\varphi & -\sin\theta.\cos\varphi & -\cos\theta.\sin\varphi & \cos\theta.\cos\varphi \end{pmatrix}$$

To calculate rotational invariance (matrix **R**), make θ = φ. To calculate twist invariance (matrix **T**), make θ = -φ.

The projection on the "transmitted" port of an analyzer acting in the a-subspace and oriented at angle α is represented by the matrix **Qa(α)**:

$$\begin{pmatrix} \cos^2\alpha & 0 & \cos\alpha.\sin\alpha & 0 \\ 0 & \cos^2\alpha & 0 & \cos\alpha.\sin\alpha \\ \cos\alpha.\sin\alpha & 0 & \sin^2\alpha & 0 \\ 0 & \cos\alpha.\sin\alpha & 0 & \sin^2\alpha \end{pmatrix}$$

and the same for an analyzer acting in the b-subspace oriented at angle α is represented by the matrix **Qb(α)**:

$$\begin{pmatrix} \cos^2\alpha & \cos\alpha.\sin\alpha & 0 & 0 \\ \cos\alpha.\sin\alpha & \sin^2\alpha & 0 & 0 \\ 0 & 0 & \cos^2\alpha & \cos\alpha.\sin\alpha \\ 0 & 0 & \cos\alpha.\sin\alpha & \sin^2\alpha \end{pmatrix}$$

The matrix **Qa(β)⊗Qb(α)** (necessary to calculate the probability of coincidence) is the product of these last two matrices, one with angle α and the other one with angle β.